\documentclass[11pt,a4paper]{article}
\usepackage{jheppub}
\usepackage[dvipsnames, svgnames, x11names]{xcolor}
\usepackage{lmodern}
\usepackage[utf8]{inputenc}
\usepackage{color}
\usepackage{float}
\usepackage{amsmath,bm}
\usepackage{amssymb}
\usepackage{graphicx}
\usepackage{setspace}
\usepackage{subfigure}
\usepackage{amsthm}
\usepackage{amsfonts}
\usepackage{braket}
\usepackage{multirow}
\usepackage{ulem}
\usepackage{slashed}
\usepackage{babel}
\usepackage{lipsum}
\usepackage{hyperref}
\usepackage{url}
\hypersetup{
	colorlinks=true,
	linkcolor=cyan,
	urlcolor=blue,
	citecolor=red,
}

\title{Spontaneous Scalarization of Brane Black Holes:\\ Quantum-Enhanced Tachyonic Instabilities}

\author[a]{Yuxuan Liu}
\author[b,c]{Yi Ling}
\author[d]{Qian Chen}

\affiliation[a]{
Institute of Quantum Physics, School of Physics, Central South University, Changsha 410083, China
}
\affiliation[b]{
 Institute of High Energy Physics, Chinese Academy of Sciences, Beijing 100049, China
}
\affiliation[c]{
School of Physics, University of Chinese Academy of Sciences, Beijing 100049, China
}
\affiliation[d]{
School of Physics, Xi'an Jiaotong University, Xi'an 710049, China
}

\emailAdd{liuyuxuan93@csu.edu.cn}
\emailAdd{lingy@ihep.ac.cn}
\emailAdd{chenqian192@mails.ucas.ac.cn}

\abstract{
We investigate the spontaneous scalarization of brane-localized charged black holes, focusing on the role of quantum-enhanced tachyonic instabilities. By solving the coupled bulk-brane equations numerically within the Einstein-DeTurck formulation, we construct fully backreacted static black hole solutions and map the system from the semi-classical limit to the strongly coupled quantum-dominated regime. We demonstrate that the holographic quantum effects, parameterized by $\kappa$, significantly modify the background geometry. Stability analysis, conducted via both effective potential diagnostics and quasinormal mode calculations, confirms that this geometric deformation deepens the negative well of the effective potential, thereby triggering a tachyonic instability. Integrating these results, we establish the global phase diagram of spontaneous scalarization in the $T_h/\mu-\kappa$ parameter space. The results reveal that these quantum-enhanced instabilities promote scalarization by substantially raising the critical temperature, providing a clear quantitative signature of quantum effects on black hole hair formation.
}

\begin{document}

\maketitle

% ===================================================================
\section{Introduction}
\label{sec:intro}
% ===================================================================
Within classical general relativity, the properties of stationary black holes in electrovacuum are tightly constrained by the uniqueness and no-hair theorems~\cite{Chrusciel:2012jk}. Consequently, the existence of extra degrees of freedom, such as scalar hair, represents a primary signature indicating deviations from standard general relativity \cite{Volkov:1998cc,Barausse:2012da,Cardoso:2016ryw,Sotiriou:2015pka}.

A prominent mechanism for generating hairy black holes is spontaneous scalarization. Originally formulated in the context of neutron stars \cite{Damour:1993hw}, this phenomenon allows a compact object to spontaneously develop a nontrivial scalar profile when specific parameters cross a critical threshold. For black holes, this dynamical transition is primarily driven by a tachyonic instability. In classical setups, evading the stability bounds necessitates explicit interaction channels—such as non-minimal couplings between the scalar field and spacetime curvature invariants~\cite{Antoniou:2017acq,Doneva:2017bvd,Dima:2020yac,East:2021bqk,Guo:2020zqm} or matter source terms like the Maxwell invariant~\cite{Herdeiro:2018wub,Doneva:2018rou,Zhang:2021etr,Hartnoll:2008vx,Ling:2020qdd}. These couplings provide a negative contribution to the effective mass squared of the scalar field, excavating a potential well in the near-horizon region that induces scalar condensation. Without such non-minimal couplings, classical stability bounds rigorously forbid the system from developing tachyonic instabilities in standard geometries~\cite{Ishibashi:2004wx,Hertog:2006rr,PhysRevD.5.1239,Bekenstein:1995un,Herdeiro:2015waa}.

Among the theoretical frameworks supporting scalarization, the Einstein-Maxwell-scalar (EMs) theory provides a well-posed system that allows for the study of fully nonlinear dynamics without the pathological issues often encountered in higher-curvature theories \cite{East:2021bqk, Ripley:2019hxt,Herdeiro:2018wub, Bosch:2016vcp, Zhang:2021etr}. In asymptotically anti-de Sitter (AdS) spacetimes, the dynamics of charged black holes in the EMs theory become particularly rich. The confining nature of the AdS boundary, combined with the near-horizon AdS$_2$ geometry of near-extremal solutions, facilitates instabilities when the effective mass of the scalar perturbation violates the Breitenlohner-Freedman (BF) bound~\cite{breitenlohner1982stability}. The resulting phase space exhibits complex structures, often supporting branches of scalarized solutions that are thermodynamically preferred over the vacuum Reissner-Nordström-AdS (RN-AdS) counterparts \cite{Hartnoll:2008vx, Zhang:2021etr, Chen:2023eru}.

While the scalarization mechanisms discussed above rely on classical non-minimal couplings, quantum effects introduce fundamental modifications to black hole dynamics and horizon structures~\cite{hawking1974black,Page:1993wv,Anderson:1993ni,Casals:2016odj,Ori:2025zhe}. A systematic framework to investigate these macroscopic quantum phenomena is braneworld holography~\cite{Karch:2000ct,Karch:2000gx,Takayanagi:2011zk}. In this configuration, a $d$-dimensional Planck brane is embedded in a $(d+1)$-dimensional asymptotically AdS bulk. By integrating out the ultraviolet degrees of freedom, an effective $d$-dimensional gravity theory is induced on the brane, naturally coupled to a conformal field theory (CFT) matter sector with a large central charge~\cite{Emparan:2002px,Emparan:2020znc,Chen:2020uac,Almheiri:2019hni,Geng:2020qvw} (for recent progress, see~\cite{Geng:2026wok}). The competition between the classical gravity sector and the quantum matter sector is parameterized by the brane tension, providing a controlled framework to explore the transition from semi-classical gravity to a strongly coupled quantum-dominated regime.

In this paper, we investigate the spontaneous scalarization of brane-localized charged black holes, focusing specifically on how the quantum effect enhances tachyonic instabilities. By numerically solving the coupled bulk-brane equations within the Einstein-DeTurck formulation, we construct fully backreacted static black hole solutions, encompassing both the bald charged black hole branch and the scalarized hairy branch. We evaluate the linear stability of the bald background by deriving the effective potential and calculating the quasi-normal modes (QNMs). The analysis demonstrates that the quantum-induced geometric deformation strictly deepens the negative well of the effective potential. This structural modification makes the system highly susceptible to tachyonic instability, thereby promoting spontaneous scalarization and substantially raising the critical phase transition temperature compared to the pure classical regime.

The paper is organized as follows. 
In section \ref{sec:model}, we establish the theoretical framework by defining the five-dimensional bulk action and the brane-localized matter sector, and we outline the Einstein-DeTurck numerical methodology. 
In section \ref{sec:quantumblackholes}, we present the static numerical solutions, categorizing them into the bald and scalarized hairy branches, and verify the geometric convergence to the analytical RN-AdS$_4$ solution in the semi-classical limit. 
Section \ref{sec:stability} is devoted to the stability analysis: we utilize the effective potential to qualitatively diagnose the tachyonic instability and perform a rigorous QNM calculation to determine the instability onset, ultimately establishing the global phase diagram in the $T_h/\mu-\kappa$ parameter space. Finally, we summarize our findings and discuss their implications in section \ref{sec:conc}.

% ===================================================================
\section{Maxwell--Scalar Black Holes on the brane}
\label{sec:model}
% ===================================================================
In this section, we establish the theoretical framework and numerical methodology for constructing Maxwell-scalar black holes localized on a Planck brane. First, we define the five-dimensional bulk action and the brane-localized matter sector, deriving the vacuum Einstein equations and the Israel junction conditions. We then introduce the static geometric ansatz within the Einstein-DeTurck formulation, specify the complete boundary conditions, and outline the pseudo-spectral method employed to solve the resulting coupled differential equations.

\subsection{Bulk geometry and brane embedding}

We consider a four-dimensional Maxwell-scalar (MS) system localized on a Planck brane $\mathcal{B}$ embedded in a five-dimensional asymptotically AdS bulk. The total action is
\begin{align}\label{eq:Action}
I=&\frac{1}{16\pi G_N} \Bigg[ \int_\mathcal{M} d^{5}x
\sqrt{-g}\left(R+\frac{12}{L^2}\right)+2\int_{\partial}d^{4}x\sqrt{-h_{\partial}}K_{\partial}\nonumber\\
&+2\int_{\mathcal{B}}d^{4}x\sqrt{-h_{\mathcal{B}}}K_{\mathcal{B}}\Bigg] - \int_\mathcal{B} d^4 x
\sqrt{-h}\mathcal{L}_M + \text{Junction terms on $\partial\mathcal{B}$}.
\end{align}
where $G_N$ is the bulk Newton constant and $L$ is the AdS radius. The tensors $h$ and $K$ denote the induced metric and the extrinsic curvature on the respective hypersurfaces, with $\partial$ representing the conformal boundary. The Lagrangian density $\mathcal{L}_M$ governs the brane-localized matter fields and is given by
\begin{equation} 
    \mathcal{L}_M
    =\sigma+\frac{1}{4}\mathcal{F}(\Phi)F_{ab}F^{ab}
    +\nabla_a\Phi \nabla^a\Phi
    +m^2 \Phi^2,
\end{equation}
where $\sigma$ is the brane tension, $F=dA$ is the Maxwell field strength, and the Latin index $a$ runs over the brane coordinates. The real scalar field $\Phi$ is non-minimally coupled to the gauge field via the coupling function $\mathcal{F}(\Phi)$.

Since the matter fields are strictly localized on the brane, the bulk geometry satisfies the vacuum Einstein equations:
\begin{equation}\label{eq:eomofbulkmetric}
    R_{\mu\nu}-\frac{1}{2}R g_{\mu\nu}-\frac{6}{L^2}g_{\mu\nu}=0.
\end{equation}
The brane embedding is determined by the Neumann boundary conditions (the Israel junction conditions):
\begin{align}\label{eq:neumanncondition}
    K_{ab}-K h_{ab}= 8 \pi G_N T_{ab},
\end{align}
where the stress-energy tensor of the localized matter reads
\begin{align}
    T_{ab}
    :=\frac{2}{\sqrt{-h}}\frac{\delta}{\delta h_{ab}}
    \left(\sqrt{-h}\mathcal{L}_M\right)
    =-h_{ab}\mathcal{L}_M + \mathcal{F}(\Phi)F_a^{\ c} F_{bc}
    +2\nabla_a\Phi \nabla_b \Phi .
\end{align}
Varying the matter action yields the equations of motion for the scalar and gauge fields:
\begin{align}
    \nabla^a \nabla_a \Phi &= \frac{1}{8} \frac{d \mathcal{F}(\Phi)}{d \Phi}F_{ab}F^{ab}+m^2 \Phi, \label{eq:eomofscalar}\\
    \nabla_a [\mathcal{F}(\Phi)F^{ab}] &= 0 . \label{eq:eomofgauge}
\end{align}

To ensure the existence of a bald charged black hole branch and preserve the $\mathbb{Z}_2$ symmetry ($\Phi \to -\Phi$) of the scalar field, we choose the coupling function
\begin{equation}
    \mathcal{F}(\Phi)=1+\alpha \Phi^2,
\end{equation}
where the coupling constant $\alpha$ is strictly positive.

\subsection{Ansatz for numerics}
The system involves bulk partial differential equations (PDEs) in two variables coupled to ordinary differential equations (ODEs) on the brane. We solve this coupled system numerically.

Instead of solving the vacuum Einstein equations \eqref{eq:eomofbulkmetric} directly, we employ the Einstein-DeTurck formulation \cite{Dias:2015nua} and solve the modified equations
\begin{equation}\label{eq:eomofdeturck}
    R_{\mu\nu}+4R g_{\mu\nu}-\nabla_{(\mu}\xi_{\nu)}=0,
\end{equation}
where the DeTurck vector is defined as $\xi^{\mu}:=\left[\Gamma_{\nu \sigma}^{\mu}(g)-\Gamma_{\nu \sigma}^{\mu}(\bar{g})\right] g^{\nu \sigma}$, and $\bar{g}$ is the reference metric. At Dirichlet boundaries, $\bar{g}$ is chosen to satisfy the same boundary conditions as $g$. We implement the Einstein--DeTurck method for the bulk gravitational equations following Ref.~\cite{Almheiri:2019psy} and solve them together with the dynamical Maxwell and scalar equations localized on the brane as a coupled bulk--brane system.

We start with the planar Schwarzschild-AdS$_{5}$ geometry
\begin{align}\label{eq:SAdS}
  ds^2=&\frac{L^2}{z^2}\left[-f(z)dt^2+\frac{dz^2}{f(z)}+\sum_{i=1}^{3}dw_i^2\right],\\
  f(z)=&1-\frac{z^4}{z_h^4},
\end{align}
where $w_i$ are the spatial coordinates and the conformal boundary is located at $z=0$. Let $\theta$ be the dihedral angle between the brane and the conformal boundary at their intersection. If the brane is embedded at
\begin{align}\label{eq:branelocation} 
   w_1 + \cot \theta z =0,
\end{align}
the Neumann boundary conditions \eqref{eq:neumanncondition} will force the bulk geometry to deviate significantly from \eqref{eq:SAdS} deep in the bulk \cite{Randall:1999vf,Dvali:2000hr,Karch:2000ct,Takayanagi:2011zk}.

For numerical convenience, we introduce the compactified coordinates $x, y \in (0,1)$ defined by
\begin{equation}
    \frac{x}{1-x}:=w_1 + \cot \theta z \text{ , and } y:= \sqrt{1-\frac{z}{z_h}}.
\end{equation}
In these coordinates, the brane is located at $x=0$, spatial infinity where the geometry approaches planar Schwarzschild-AdS$_{5}$ is at $x=1$, the event horizon is at $y=0$, and the conformal boundary is at $y=1$.

Since the brane backreaction breaks the translational symmetry along the $x$ direction, the most general static ansatz for the bulk metric is
\begin{align}\label{eq:detuckansatz}
\mathrm{d} s^2=&\frac{L^2}{\left(1-y^2\right)^2}\left\{-y^2 \mathcal{P}(y) y_h^2 Q_1 \mathrm{~d} t^2+\frac{4 Q_2 \mathrm{~d} y^2}{\mathcal{P}(y)}+\right. \\ \nonumber
&\left.\frac{Q_4}{(1-x)^4}\left[y_h \mathrm{d} x+2(1-x)^2 y Q_3 \mathrm{~d} y\right]^2+y_h^2 Q_5\left(\mathrm{~d} w_2^2+\mathrm{d} w_3^2\right)\right\},
\end{align}
where $y_h:=1/z_h$,
\begin{equation}
    \mathcal{P}(y):=(2-y^2)(2-2y^2+y^4),
\end{equation}
 and the unknown functions $Q_i(x,y)$ ($i=1,\dots,5$) govern the bulk geometry.
The reference metric $\bar{g}$ is obtained by setting $Q_1=Q_2=Q_4=Q_5=1$ and $Q_3=\cot\theta$.

For the brane-localized matter fields at $x=0$, we assume the static ansatz
\begin{equation}
    \Phi=\phi(y) \text{, and , } A=A(y)dt.
\end{equation}
Consequently, the system comprises seven unknown functions: the five bulk metric components $Q_i(x,y)$ and the two brane matter fields $\phi(y)$ and $A(y)$, where the latter couple to $Q_i$ exclusively at the brane boundary $x=0$.

We impose the boundary conditions as follows. At the horizon ($y=0$), regularity requires
\begin{equation}\label{eq:bcsx0}
    \partial_y Q_i =0 , \; \phi'(0)=0, \; \text{ and } \; A(0)=0,
\end{equation}
along with $Q_1(x,0)=Q_2(x,0)$ and $A'(0)=0$, which fixes the Hawking temperature to
\begin{equation}\label{eq:temperature}
    T_h=\frac{y_h}{\pi}.
\end{equation}
At the conformal boundary ($y=1$), the geometry asymptotes to \eqref{eq:SAdS}, demanding $Q_1=Q_2=Q_4=Q_5=1$ and $Q_3=\cot\theta$. This AdS asymptotic behavior fixes the brane tension and the scalar mass:
\begin{equation}
    \sigma = \frac{6\cos \theta}{L}, \; \text{ and } \; m^2=\frac{-2\sin^2\theta}{L^2}.
\end{equation}
Near the conformal boundary ($z\to0$), the matter fields behave as
\begin{equation}
    \phi(z)\bigg|_{z=0} = \phi_s z +\phi_v z^2 +\cdots, \text{ and } A(z)\bigg|_{z=0} = \mu -\rho z  +\cdots,
\end{equation}
where $\phi_s$ is the scalar source, $\phi_v$ is the vacuum expectation value (VEV), $\mu$ is the chemical potential, and $\rho$ is the charge density.
At the brane ($x=0$), we impose the Neumann boundary conditions \eqref{eq:neumanncondition}, along with the gauge-fixing condition $\xi_a n^a=0$ (where $n^a$ is the unit normal to the brane) and the geometric requirement $Q_3(0,y)=\cot\theta$.
The matter equations \eqref{eq:eomofscalar} and \eqref{eq:eomofgauge} are evaluated exclusively at $x=0$.
At spatial infinity ($x=1$), the brane backreaction vanishes, so the metric returns to the planar Schwarzschild-AdS$_5$ form, enforcing $Q_1=Q_2=Q_4=Q_5=1$ and $Q_3=\cot\theta$.

We discretize the coupled PDE-ODE system using a pseudo-spectral collocation method on a Chebyshev-Gauss-Lobatto grid and solve the resulting non-linear algebraic equations via the Newton-Raphson method.

A crucial step in the Einstein-DeTurck approach is ensuring that the numerical result represents a genuine solution to the Einstein equations ($\xi^{\mu} = 0$) rather than a DeTurck soliton ($\xi^{\mu} \neq 0$). While the non-existence of DeTurck solitons is mathematically proven for static geometries with Dirichlet boundaries, a rigorous proof is currently lacking for Neumann boundaries \cite{Almheiri:2019psy,Figueras:2011va,Figueras:2016nmo}. Nevertheless, since the boundary value problem is strongly elliptic, solutions are locally unique, meaning any DeTurck soliton would be isolated from the physical solution. In practice, we monitor $\xi^2:=\xi_\mu\xi^\mu$ and verify that its maximum absolute value decreases as the collocation grid is refined \cite{Liu:2022pan}. A representative convergence test and the spatial distribution of $\xi^2$ are presented in Appendix~\ref{app:deturck-convergence}.

% ===================================================================
\section{Static solutions}
\label{sec:quantumblackholes}
% ===================================================================
In this section, we present the numerical static solutions of the coupled system, categorizing them into bald and hairy branches. We first formulate the four-dimensional effective theory on the brane and introduce the parameter $\kappa$ to quantify the strength of the holographic quantum effects. Next, we construct the bald charged black hole solutions, explicitly verifying their geometric convergence to the exact planar RN-AdS$_4$ metric in the semi-classical limit. Finally, we investigate the scalarized branch, demonstrating how the quantum effects, alongside the non-minimal coupling and temperature, governs the strength of the scalar condensation near the horizon.

\subsection{Quantum matter sector on the brane}

Evaluating the bulk action \eqref{eq:Action} on-shell yields a four-dimensional effective theory on the brane~\cite{Chen:2020uac,Emparan:2020znc}:
\begin{align}\label{eq:braneaction}
    I_{\rm eff}=&\frac{1}{16\pi G_n}\int_\mathcal{B} d^4x \sqrt{-h}\left(\frac{6}{l^2}+R_h\right)+\mathcal{O}(R_h^2)+\cdots\nonumber\\
    &-\int_{\mathcal{B}}d^{4}x\sqrt{-h}\left(\frac{1}{4}\mathcal{F}(\Phi)F_{ab}F^{ab}
    +\nabla_a\Phi \nabla^a\Phi
    +m^2 \Phi^2\right) + I_{\rm CFT},
\end{align}
where the effective 4D Newton constant $G_n$ and the AdS$_4$ radius $l$ are given by
\begin{align}\label{eq:parameterrelation}
      G_n = \frac{2G_N}{L},\quad\text{and}\quad l = \frac{L}{\sqrt{2(1- \cos \theta)}}.
\end{align}
The action $I_{\rm CFT}$ represents the dual conformal field theory (CFT) matter sector on the brane~\cite{Emparan:1999pm,Emparan:2002px,Emparan:2020znc}, characterized by a large central charge $c \sim L^3/G_N \gg 1$.

The competition between the classical gravity sector and the quantum CFT sector is governed by the dimensionless quantum effect parameter:
\begin{equation}
    \kappa := \frac{1}{1- \cos \theta}\sim\frac{l^2/G_n}{c}.
\end{equation}
In the semi-classical limit ($\kappa\gg1$), the parameters naturally exhibit the hierarchy~\cite{Almheiri:2019hni}:
\begin{equation}
    l^2/G_n \gg c \gg 1.
\end{equation}
Conversely, in the strongly coupled quantum-dominated regime $(\kappa \sim \mathcal{O}(1))$, the hierarchy becomes:
\begin{equation}
    c \sim l^2/G_n \gg 1.
\end{equation}
For numerical convenience, we set $\{L, 8\pi G_N, \mu\}=\{1, 1, 1\}$ throughout this work, which numerically ensures that $c\sim L^3/G_N = 8\pi \gg 1$.

\begin{figure}
  \centering
    \subfigure[]{\label{fig:properdistance1}
  \includegraphics[height=0.4\linewidth]{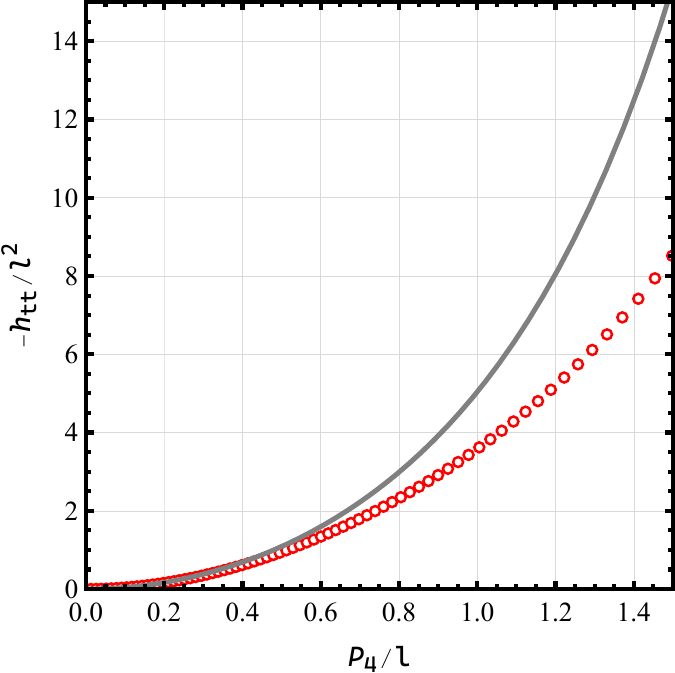}}
    \hspace{0pt}
    \subfigure[]{\label{fig:properdistance2}
  \includegraphics[height=0.4\linewidth]{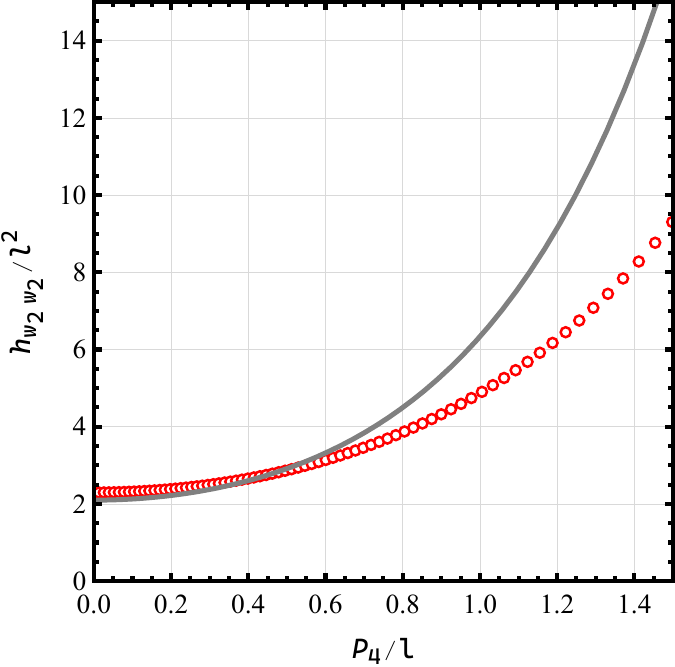}}\\
    \subfigure[]{\label{fig:properdistance3}
  \includegraphics[height=0.4\linewidth]{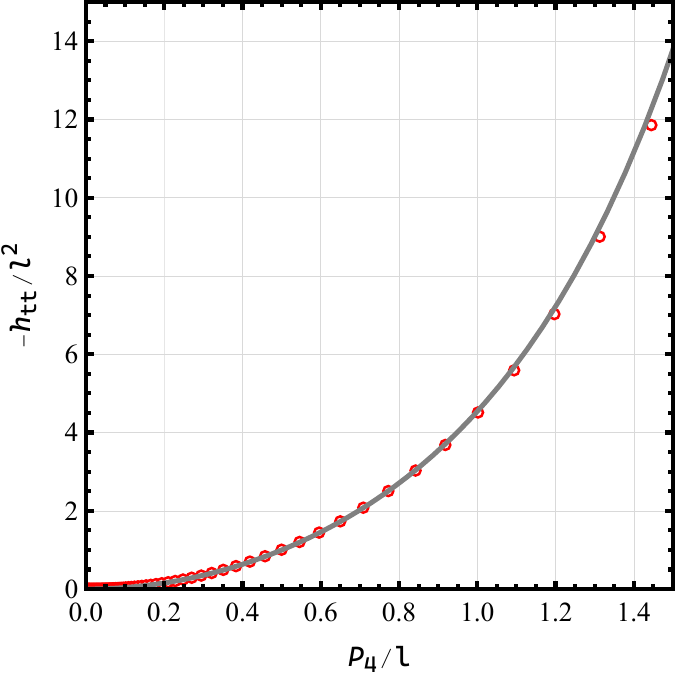}}
  \hspace{0pt}
    \subfigure[]{\label{fig:properdistance4}
  \includegraphics[height=0.4\linewidth]{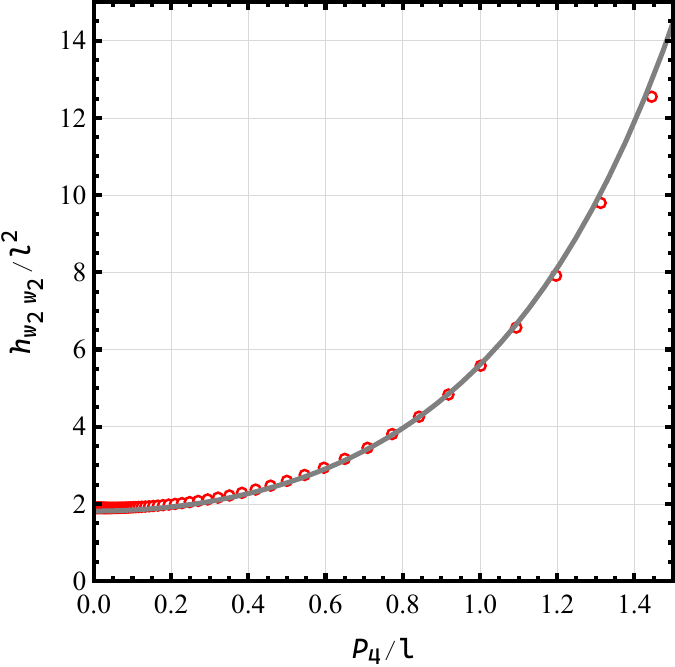}}\\
\caption{Metric components $-h_{tt} / l^2$ and $h_{w_2w_2} / l^2$ evaluated on the brane ($x=0$) as functions of the proper radial distance from the horizon, $P_{4}/l$. The top row corresponds to the quantum-dominated case with $\{Z_h,\kappa\} \approx \{0.69042, 1\}$, while the bottom row represents the semi-classical case with $\{Z_h,\kappa\} \approx \{0.74226, 8.89924\}$. Red open markers denote the numerical data, and solid gray lines represent the exact 4D planar RN-AdS black hole geometry given in \eqref{eq:rnads4}.}\label{fig:properdistance}
\end{figure}

\subsection{Bald branch}
When the scalar field $\Phi$ is turned off, the system admits a bald charged black hole solution on the brane. Since the induced metric on the brane is solved numerically, its exact geometry is generally unknown. However, we find that in the semi-classical limit ($\kappa \gg 1$), the brane geometry asymptotically approaches the metric of a four-dimensional planar RN-AdS black hole.

\begin{figure}
  \centering
    \subfigure[]{\label{fig:bald1}
  \includegraphics[height=0.4\linewidth]{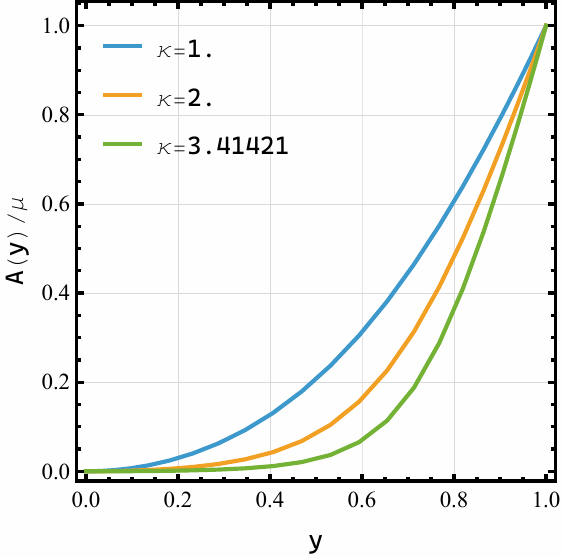}}
    \hspace{0pt}
    \subfigure[]{\label{fig:bald2}
  \includegraphics[height=0.4\linewidth]{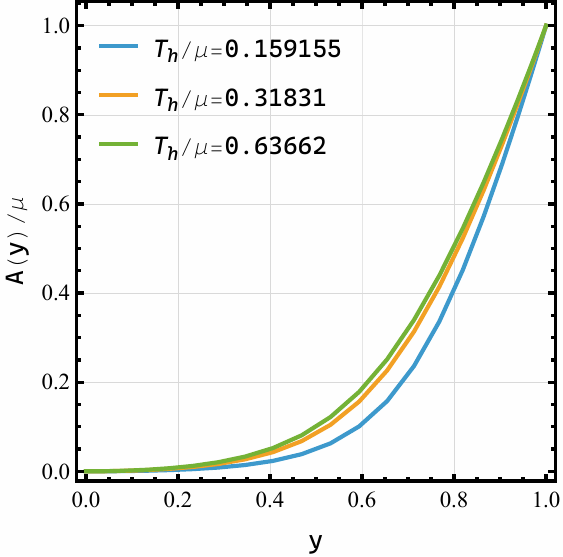}}\\
\caption{Radial profiles of the dimensionless gauge field $A(y)/\mu$ in the bald phase. Panels (a) and (b) illustrate the effects of the quantum parameter $\kappa$ and the dimensionless temperature $T_h/\mu$, respectively. In (a), the temperature is fixed at $T_h/\mu \approx 0.31831$. In (b), the quantum parameter is fixed at $\kappa\approx2$.}\label{fig:bald}
\end{figure}

To explicitly demonstrate this asymptotic behavior, we write down the standard line element of a planar RN-AdS$_4$ black hole:
\begin{equation}\label{eq:rnads4}
    ds^2_4= \frac{l^2}{Z^2}\left[-\mathcal{G}(Z)dt^2+\frac{dZ^2}{\mathcal{G}(Z)}+dw_2^2+dw_3^2\right],
\end{equation}
where the blackening factor is given by
\begin{equation}
    \mathcal{G}(Z)=1-\left(\frac{Z}{Z_h}\right)^3-\frac{\mu_4^2 Z_h^2}{4 l^2}\left(\frac{Z}{Z_h}\right)^3\left(1-\frac{Z}{Z_h}\right).
\end{equation}
Here, the outer horizon is located at $Z=Z_h$, and $\mu_4$ is the chemical potential. The associated Hawking temperature is
\begin{equation}
    T_{h4}=\frac{1}{4\pi}\left(\frac{3}{Z_h}-\frac{\mu_4^2 Z_h}{4l^2}\right).
\end{equation}
To match the thermodynamic background of our numerical setup \eqref{eq:temperature}, we set $\mu_4=\mu$ and require
\begin{equation}
    Z_h =\frac{2 \left(\sqrt{16 l^4 y_h^2+3 l^2 \mu_4^2}-4 l^2 y_h\right)}{\mu_4^2}.
\end{equation}
where $l$ is determined by \eqref{eq:parameterrelation}.

The numerical induced metric on the brane evaluated at $x=0$ reads
\begin{equation}\label{eq:inducedmetric}
\begin{aligned}
\mathrm{d} s_{\mathcal{B}}^2=\frac{L^2}{\left(1-y^2\right)^2}\left\{-y^2 \mathcal{P}(y) y_h^2 Q_1  \mathrm{d} t^2+4\left[\frac{Q_2 }{\mathcal{P}(y)}\right.\right. & \left.+y^2 \cot ^2 \theta Q_4 \right] \mathrm{d} y^2 \\
& \left.+y_h^2 Q_5 \left(\mathrm{d} w_2^2+\mathrm{d} w_3^2\right)\right\},
\end{aligned}
\end{equation}
where all metric functions $\{Q_1,Q_2,Q_4,Q_5\}$ are evaluated at the brane location $x=0$. To perform a coordinate-independent comparison between \eqref{eq:rnads4} and \eqref{eq:inducedmetric}, we evaluate the metric components $-h_{tt} / l^2$ and $h_{w_2 w_2} / l^2$ as functions of the proper radial distance from the horizon, $P_4$.

\begin{figure}
  \centering
    \subfigure[]{\label{fig:hairy1}
  \includegraphics[height=0.4\linewidth]{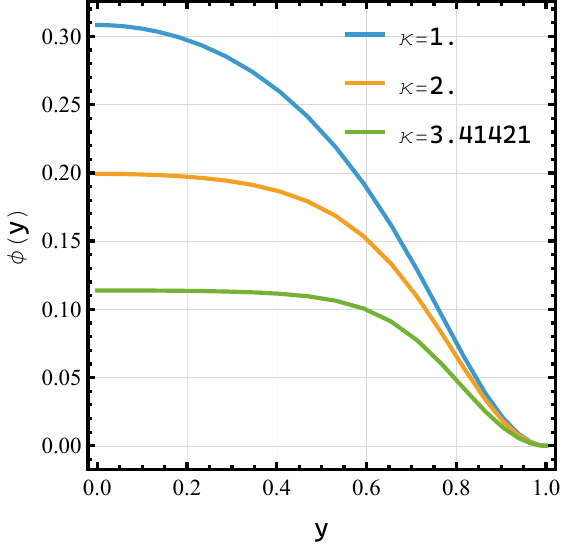}}
    \hspace{0pt}
    \subfigure[]{\label{fig:hairy2}
  \includegraphics[height=0.4\linewidth]{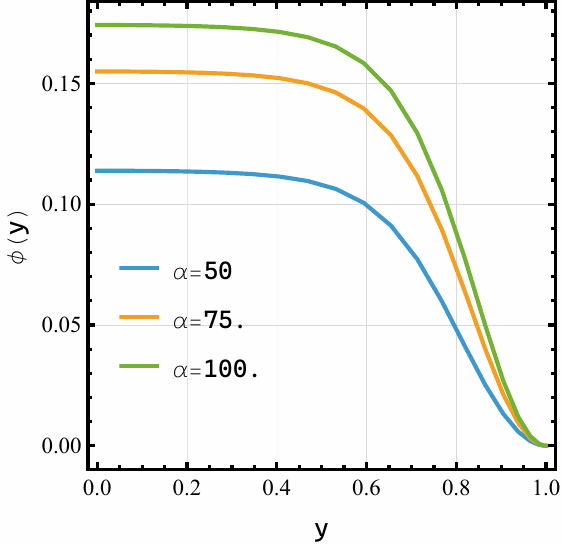}}\\
    \subfigure[]{\label{fig:hairy3}
  \includegraphics[height=0.4\linewidth]{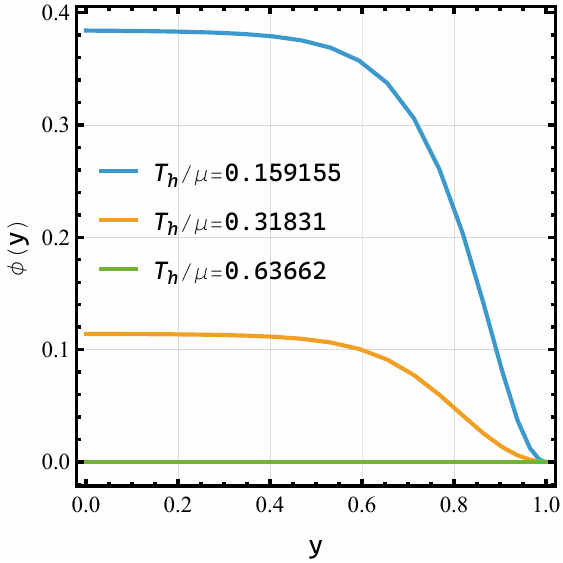}}
  \hspace{0pt}
    \subfigure[]{\label{fig:hairy4}
  \includegraphics[height=0.4\linewidth]{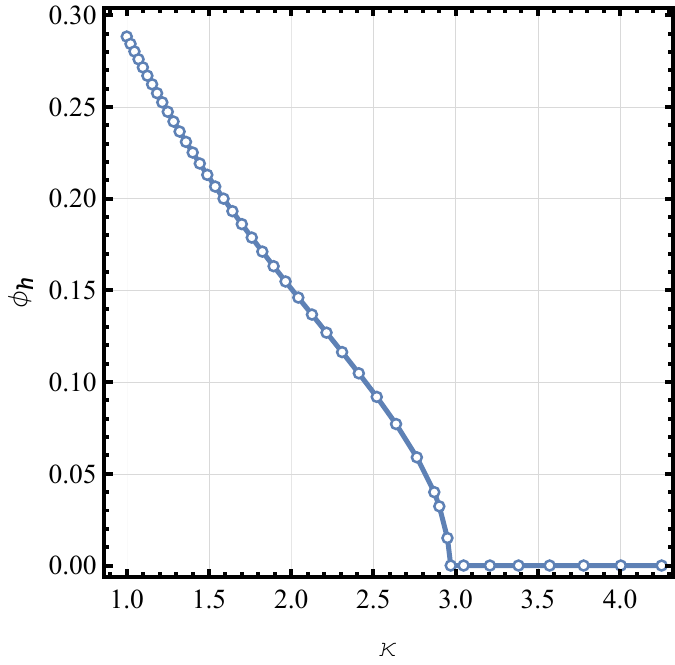}}\\
\caption{Properties of the scalarized black hole solutions. Panels (a), (b), and (c) display the radial profiles of the scalar field $\phi(y)$ illustrating the influence of $\kappa$, the non-minimal coupling constant $\alpha$, and the temperature $T_h/\mu$, respectively. Panel (d) shows the value of the scalar field at the horizon, $\phi_h:=\phi(0)$, as a function of $\kappa$. The fixed parameters are $\{\alpha,T_h/\mu\} \approx \{50,0.31831\}$ for (a), $\{\kappa,T_h/\mu\}\approx\{3.41421,0.31831\}$ for (b), and $\{\kappa, \alpha\} \approx \{1.17157, 25\}$ for (c), while $\{\alpha,T_h/\mu\}\approx\{25,0.28403\}$ is fixed for (d).}\label{fig:hairy}
\end{figure}

For the analytical RN-AdS$_4$ metric \eqref{eq:rnads4}, the proper distance is
\begin{equation}
    \frac{P_4(Z)}{l}=\int^{Z_h}_Z \frac{dZ}{Z}\sqrt{\frac{1}{\mathcal{G}(Z)}},
\end{equation}
while for the numerical induced metric \eqref{eq:inducedmetric}, it is given by
\begin{equation}
    \frac{P_4(y)}{l}=\frac{2L}{l}\int_0^y\frac{dy}{(1-y)^2}\sqrt{\frac{Q_2}{\mathcal{P}(y)}+y^2 \cot^2\theta Q_4}.
\end{equation}
As illustrated in Figure~\ref{fig:properdistance}, the numerical metric components precisely converge to the analytical 4D planar RN-AdS profiles in the limit $\kappa \gg 1$.

Furthermore, the radial profiles of the gauge field in the bald phase are presented in Figure~\ref{fig:bald}. The results indicate that increasing the quantum parameter $\kappa$ or lowering the dimensionless temperature $T_h/\mu$ pushes the electric field configuration towards the conformal boundary ($y \to 1$).

\subsection{Hairy branch}
In specific parameter regimes, the bald black hole becomes unstable to scalar perturbations, leading to the emergence of a charged hairy black hole branch on the brane. The properties of these scalarized solutions are strongly dictated by the quantum effects. As illustrated in Figure~\ref{fig:hairy1}, the scalar condensation near the horizon becomes significantly more pronounced as the system enters the quantum-dominated regime (decreasing $\kappa$). This monotonic enhancement of the scalar hair by the quantum effect is explicitly depicted in Figure~\ref{fig:hairy4}.

Similarly, increasing the non-minimal coupling constant $\alpha$ (Figure~\ref{fig:hairy2}) or lowering the dimensionless temperature $T_h/\mu$ (Figure~\ref{fig:hairy3}) enhances the condensation of the scalar field. These parametric dependencies are consistent with standard spontaneous scalarization mechanisms~\cite{Hartnoll:2008vx,Ling:2020qdd,Antoniou:2017acq,Doneva:2017bvd,Dima:2020yac,East:2021bqk,Guo:2020zqm}, where stronger couplings or lower temperatures efficiently trigger the underlying tachyonic instability.

% ===================================================================
\section{Stability}
\label{sec:stability}
In this section, we investigate the stability of the bald black hole solutions against scalar perturbations. We first derive the effective potential from the linearized Klein-Gordon equation to qualitatively diagnose the onset of tachyonic instability. Subsequently, we determine the instability boundaries numerically by computing the QNMs of the scalar field. Finally, integrating these perturbative results, we construct the complete thermodynamic phase diagram of the system to illustrate how the holographic quantum effects govern the spontaneous scalarization.

% ===================================================================
\subsection{Effective potential}
The stability of the system can be preliminarily diagnosed by examining the effective potential of the scalar perturbations. A necessary condition for the onset of tachyonic instability is the presence of a sufficiently deep negative well in this effective potential.

\begin{figure}
  \centering
    \subfigure[]{\label{fig:potential1}
  \includegraphics[height=0.4\linewidth]{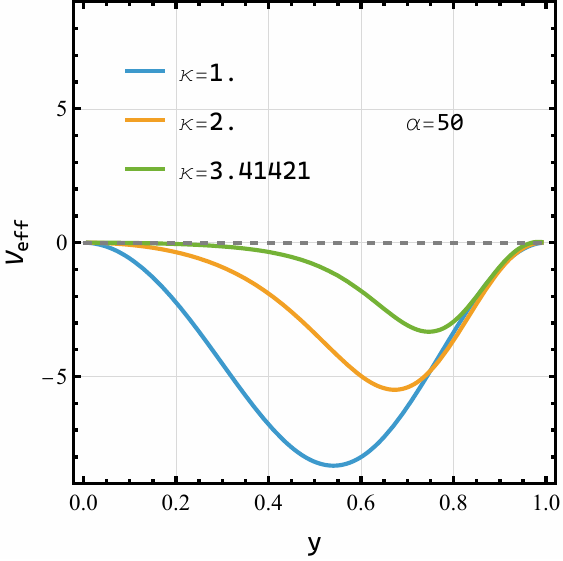}}
    \hspace{0pt}
    \subfigure[]{\label{fig:potential2}
  \includegraphics[height=0.4\linewidth]{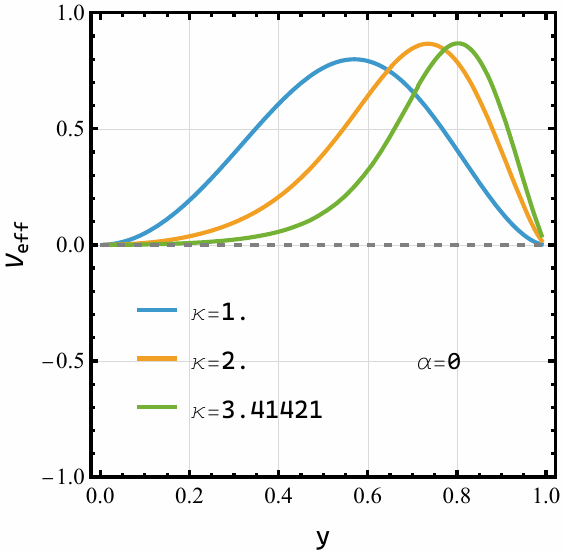}}\\
    \subfigure[]{\label{fig:potential3}
  \includegraphics[height=0.4\linewidth]{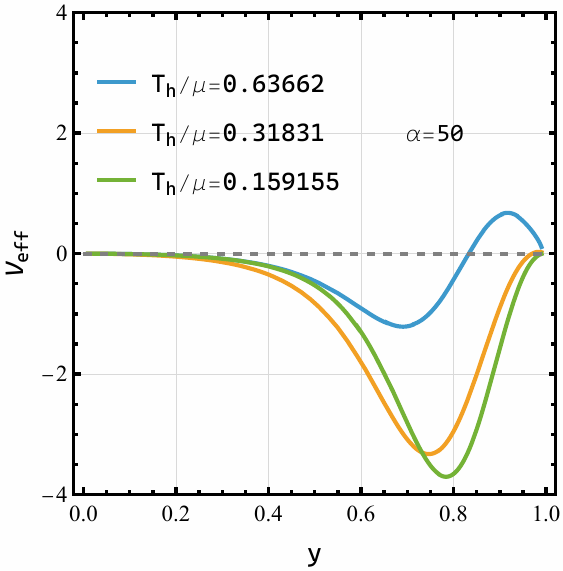}}
  \hspace{0pt}
    \subfigure[]{\label{fig:potential4}
  \includegraphics[height=0.4\linewidth]{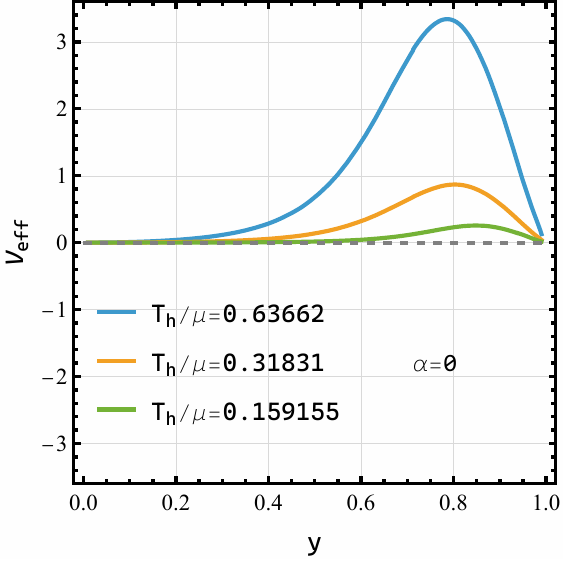}}\\
\caption{The effective potential of the bald black holes. The top row illustrates the influence of the quantum effects $\kappa$ with a fixed temperature $T_h/\mu=0.31831$. The bottom row illustrates the effect of the temperature $T_h/\mu$ with a fixed $\kappa=3.41421$. The left column shows the cases with a non-minimal coupling $\alpha=50$ (panels a and c), while the right column represents the decoupled cases with $\alpha=0$ (panels b and d).}\label{fig:potential}
\end{figure}

We consider a time-dependent, planar-symmetric linear perturbation of the scalar field on the bald background ($\phi_0=0$):
\begin{equation}
\widetilde{\phi}(t, y)=\phi_0(y)+\delta \phi(y) e^{-i \omega t} .
\end{equation}
At linear order, the scalar perturbation decouples from the metric and gauge field fluctuations. Thus, it satisfies the linearized Klein-Gordon equation \eqref{eq:eomofscalar}:
\begin{equation}\label{eq:perturbation}
    \nabla^a \nabla_a \widetilde{\phi} = \left(\frac{1}{4} \alpha F_{ab}F^{ab}+m^2\right) \widetilde{\phi}.
\end{equation}
Expanding \eqref{eq:perturbation} yields a standard second-order ordinary differential equation:
\begin{equation}
    \delta\phi''+c_1(y)\delta\phi'+c_2(y)\delta\phi=0.
\end{equation}
By introducing the generalized tortoise coordinate $r^*$, defined by
\begin{equation}
    c_3(y):=\frac{dr^*}{dy}=\frac{2 \sqrt{y^2 \mathcal{P} Q_3(0,y)^2 Q_4(0,y)+Q_2(0,y)}}{y \;y_h \mathcal{P} \sqrt{Q_1(0,y)}}
\end{equation}
and redefining the radial function as
\begin{equation}
    \delta\phi(y)= \frac{e^{-\frac{1}{2}\int c_1(y)dy}}{\sqrt{c_3(y)}} \mathcal{U}(r^*),
\end{equation}
we can recast the scalar perturbation equation \eqref{eq:perturbation} into a Schr\"odinger-like form:
\begin{equation}\label{eq:scalarperturbation}
    \frac{d^2}{d r_*^2} \mathcal{U}(r^*)+\omega^2 \mathcal{U}(r^*)-V_{\mathrm{eff}} \;\mathcal{U}(r^*) = 0.
\end{equation}
The effective potential $V_{\text{eff}}$ is explicitly given by
\begin{align}\label{eq:potential}
   V_{\text{eff}} = \frac{-\mathcal{P}}{16 F_2^2 k^2 Q_5^2} \left[ 2 Q_5 F_2 \cdot \mathcal{A}(y) + y y_h^2 F_5 Q_5 \cdot \mathcal{B}(y) + y^2 y_h^2 F_5^2 \mathcal{P} Q_1 F_2 \right],
\end{align}
where
\begin{align}
    k:=&-1+y^2,\nonumber\\
    F_5:=&k\, \partial_y Q_5-4yQ_5,\nonumber\\
    F_2:=&Q_2+y^2 \mathcal{P}Q_3^2Q_4,\nonumber\\
    \mathcal{A}(y) :=& \alpha k^4 Q_5 A'^2 + y^2 y_h^2 Q_1 \left[ 4 Q_5 \left( \mathcal{P} \left( 4 y^2 \cos^2\theta Q_4 - k + 5 y^2 - 1 \right) + 4 \sin^2\theta Q_2 \right) - F_5' \mathcal{P} k \right],\nonumber\\
    \mathcal{B}(y) :=& \mathcal{P} \left[ Q_1 \left( y k \left( \partial_y Q_2 - y^2 \cot^2\theta Q_4 \mathcal{P}' \right) + 2(9y^2+1)Q_2 \right) - y k \partial_y Q_1 Q_2 \right]\nonumber\\ &- 2 y k Q_1 Q_2 \mathcal{P}' + y^3 \mathcal{P}^2 Q_4 \cot^2\theta \left[ Q_1 \left( 20 y + k \frac{\partial_y Q_4}{Q_4} \right) - k \partial_y Q_1 \right].
\end{align}

The radial profiles of the effective potential are plotted in Figure~\ref{fig:potential}. By observation, $V_{\text{eff}}$ vanishes at both the horizon and the conformal boundary. In the absence of the non-minimal coupling ($\alpha=0$), the potential is strictly positive everywhere (Figures~\ref{fig:potential2} and \ref{fig:potential4}), precluding the existence of bound states and ensuring the stability of the bald background. In this decoupled scenario, varying the quantum parameter $\kappa$ merely shifts the peak of the potential without significantly altering its magnitude, while lowering the temperature uniformly suppresses the amplitude.

Conversely, when the non-minimal coupling is introduced ($\alpha > 0$), a negative potential well develops, satisfying the necessary condition for tachyonic instability (Figures~\ref{fig:potential1} and \ref{fig:potential3}). Crucially, strengthening the quantum effects (decreasing $\kappa$) or lowering the temperature progressively deepens this negative well. The temperature dependence is standard for holographic spontaneous symmetry breaking~\cite{Hartnoll:2008vx}. More importantly, the $\kappa$ dependence explicitly demonstrates that the quantum effects geometrically deform the background, which significantly deepens the effective potential well and promotes scalarization, provided $\alpha \neq 0$ \eqref{eq:potential}.

\begin{figure}
  \centering
  \subfigure[]{\label{fig:qnm2}
  \includegraphics[width=0.31\linewidth]{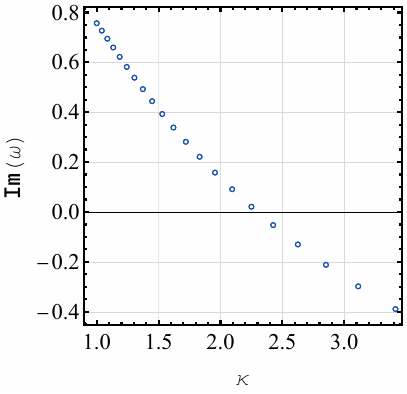}}
  \hspace{0pt}
  \subfigure[]{\label{fig:qnm4}
  \includegraphics[width=0.31\linewidth]{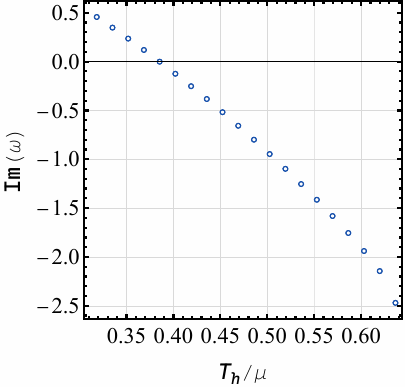}}
  \hspace{0pt}
  \subfigure[]{\label{fig:qnm6}
  \includegraphics[width=0.32\linewidth]{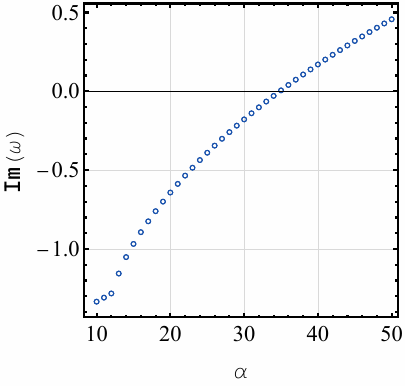}}
\caption{The imaginary part of the continuously tracked fundamental QNM as a function of (a) $\kappa$, with fixed $\{\alpha,T_h/\mu\}=\{25,0.31831\}$; (b) $T_h/\mu$, with fixed $\{\alpha,\kappa\}=\{50,3.41421\}$; and (c) $\alpha$, with fixed $\{\kappa,T_h/\mu\}=\{3.41421,0.31831\}$. The zero crossings mark the onset of the scalar instability, with $\operatorname{Im}(\omega)>0$ corresponding to a growing mode.}\label{fig:qnm}
\end{figure}

\subsection{Quasi-normal modes}
While the effective potential provides a qualitative diagnostic, a direct numerical test of instability is provided by computing the QNMs of the scalar perturbation. The bald background is unstable if there exists at least one mode with a positive imaginary part, $\text{Im}(\omega) > 0$.

Near the horizon at $y=0$, substituting $\delta\phi\sim y^\Delta$ into the perturbation equation gives the indicial equation
\begin{equation}
    4y_h^2\Delta^2+\omega^2=0,
\end{equation}
and hence $\Delta_\pm=\pm i\omega/(2y_h)$. With the time dependence $e^{-i\omega t}$, the ingoing branch is $\Delta=-i\omega/(2y_h)$. We therefore write
\begin{equation}
    \delta\phi(t,y)=e^{-i\omega t}\frac{1-y^2}{y_h}y^{-i\omega/(2y_h)}F(y),
\end{equation}
where $F(y)$ is regular on the interval $0\leq y\leq1$. After factoring out the ingoing behavior, we impose the homogeneous horizon regularity condition $F'(0)=0$. At the conformal boundary, $z=(1-y^2)/y_h$ and $\delta\phi=\phi_s z+\phi_v z^2+\cdots$; hence the source-free condition $\phi_s=0$ is imposed by $F(1)=0$. These two homogeneous conditions replace the endpoint collocation equations. The overall normalization of the eigenfunction is arbitrary and is fixed only after solving the generalized eigenvalue problem.

Because the background static solutions are strictly numerical, we determine the complex frequencies $\omega$ using the generalized eigenvalue method~\cite{Dias:2015nua}. By applying a Chebyshev pseudo-spectral discretization to the perturbation equation \eqref{eq:scalarperturbation}, the continuous differential operator is mapped into a finite-dimensional matrix, casting the problem into a generalized algebraic eigenvalue problem where $\omega$ is the corresponding eigenvalue.

Figure~\ref{fig:qnm} displays the imaginary part of the continuously tracked fundamental QNM directly. The bald black hole is stable at sufficiently high temperature or sufficiently small non-minimal coupling $\alpha$. Increasing $\alpha$, lowering $T_h/\mu$, or decreasing $\kappa$ drives $\operatorname{Im}(\omega)$ through zero and into the upper half-plane. In particular, the $\kappa$ dependence shows quantitatively that stronger holographic quantum effects promote the tachyonic instability that triggers spontaneous scalarization.

A convergence test of the QNM calculation is presented in Appendix~\ref{app:qnm-convergence}.

% ===================================================================
\subsection{Phase Diagram}
\label{sec:phasediagram}

\begin{figure}
  \centering
  \includegraphics[height=0.45\linewidth]{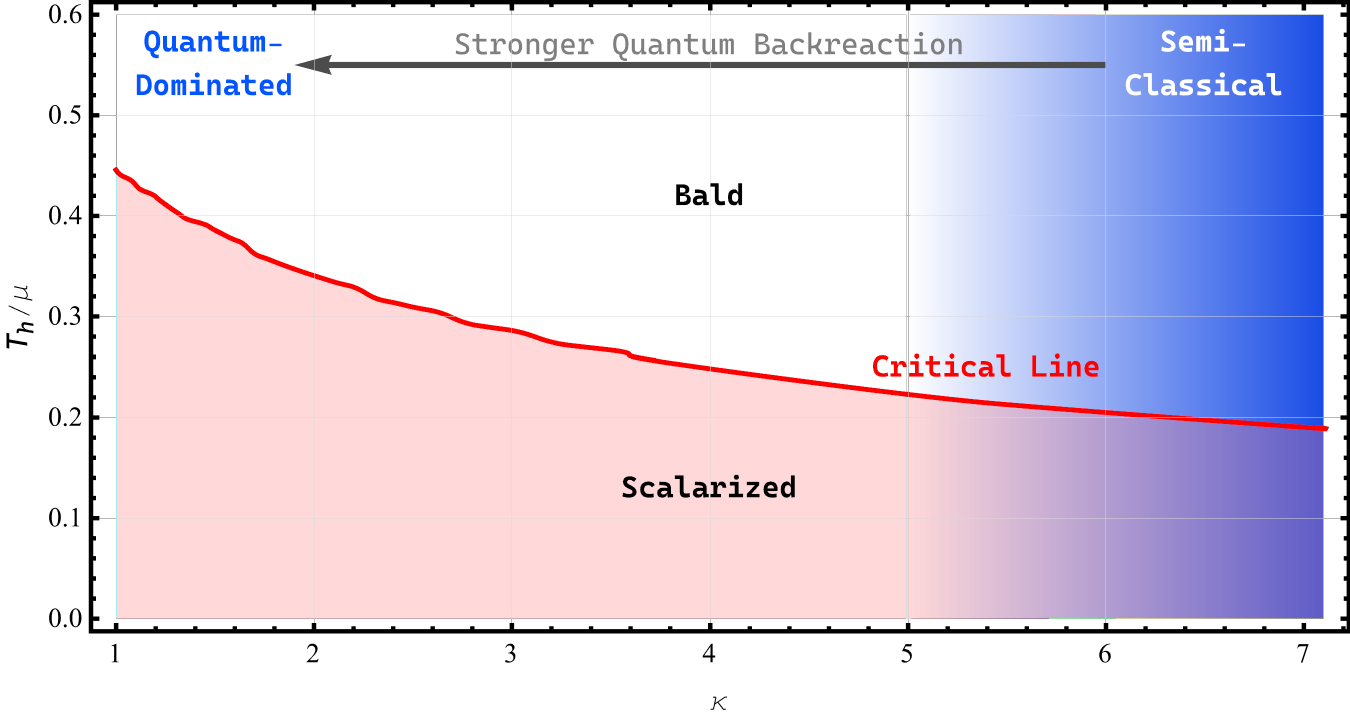}
\caption{The phase diagram of spontaneous scalarization in the $T_h/\mu-\kappa$ parameter space. The red solid line represents the critical boundary separating the bald and scalarized phases. The blue region on the right corresponds to the semi-classical limit (large $\kappa$), where the high-temperature and low-temperature regimes approach the classical bald RN-AdS$_4$ and scalarized RN-AdS$_4$ geometries, respectively. The left side corresponds to the quantum-dominated regime, demonstrating that stronger quantum effects (smaller $\kappa$) raise the critical temperature for scalarization. In this diagram, the non-minimal coupling is fixed at $\alpha=25$.}\label{fig:pd}
\end{figure}

Following the analytical derivation of the effective potential $V_{\text{eff}}$ and the perturbative QNM calculations, this section presents the overall phase diagram in the $T_h/\mu-\kappa$ parameter space (Figure~\ref{fig:pd}). This diagram comprehensively illustrates how the strength of the quantum effects, parameterized by $\kappa$, influences the onset of spontaneous scalarization.

For each fixed temperature, the critical value of $\kappa$ is determined by following the source-free scalarized branch and locating the limit in which the horizon scalar field approaches zero, $\phi_h\to0$. Repeating this procedure at different temperatures yields the critical curve shown in Figure~\ref{fig:pd}. For the tachyonic instability considered here, this bifurcation criterion is equivalent to the appearance of a normalizable static scalar zero mode: the relevant purely imaginary QNM branch reaches $\omega=0$ and $\operatorname{Im}(\omega)$ changes sign.

The core result revealed by the phase diagram is that the holographic quantum effects significantly promote spontaneous scalarization. As shown in the figure, the critical temperature curve (red solid line) separating the bald and scalarized phases increases monotonically as $\kappa$ decreases. Physically, as analyzed in previous sections, the enhancement of quantum effects modifies the background geometry, which consequently deepens the negative well of the effective potential. This geometrical effect makes the system more susceptible to tachyonic instability, allowing black holes to develop scalar hair at considerably higher temperatures compared to the classical case.

Furthermore, the phase diagram accurately reflects the expected asymptotic behaviors in different geometric limits. On the right side of the diagram in blue (large $\kappa$), the quantum effect is highly suppressed, and the system reduces to the semi-classical limit. Consequently, the background geometry in the high-temperature region approaches the classical bald RN-AdS$_4$ black hole, while the low-temperature region approaches the classical scalarized RN-AdS$_4$ solution. Conversely, on the left side, the purely classical description is substantially modified, and the system enters the quantum-dominated regime where the onset of scalarization is largely enhanced by the quantum effects.

% ===================================================================
\section{Conclusions and discussions}
\label{sec:conc}
% ===================================================================
In this paper, we investigated the spontaneous scalarization of charged black holes localized on a Planck brane. The framework is based on a four-dimensional Maxwell-scalar theory embedded in a five-dimensional asymptotically AdS bulk. By employing the Einstein-DeTurck formulation and pseudo-spectral collocation methods, we numerically solved the coupled bulk-brane system to construct the fully backreacted static configurations, encompassing both the bald and hairy black hole branches.

For the bald branch, we explicitly verified that the induced geometry on the brane cleanly approaches the analytical four-dimensional planar RN-AdS$_4$ metric in the semi-classical limit ($\kappa \gg 1$). Using the parameter $\kappa$, we continuously tuned the strength of the holographic quantum effects, allowing us to scan the physical parameter space from the classical limits into the strongly coupled quantum-dominated regime.

To determine the onset of scalarization, we systematically analyzed the stability of the bald background against linear scalar perturbations. The derivation of the effective potential revealed that the non-minimal coupling $\alpha$ provides the necessary negative potential well. The subsequent QNM calculations provided direct numerical evidence for the tachyonic instability through the emergence of fundamental modes with positive imaginary parts.

Collecting these numerical and perturbative results, we constructed the complete phase diagram in the $T_h/\mu - \kappa$ parameter space. The central conclusion of this work is that the quantum effect significantly promotes the spontaneous scalarization of black holes. The strong geometric deformation induced by the quantum effects deepens the negative well of the effective potential, which effectively makes the system more susceptible to the tachyonic instability and substantially raises the critical phase transition temperature compared to the pure classical regime.

Our framework opens up several intriguing avenues for future research:
\begin{itemize}
    \item First, the braneworld holography setup provides a natural arena for investigating the black hole information paradox. It is highly compelling to compute the entanglement entropy of the Hawking radiation using the island formula and explicitly derive the Page curve for these fully backreacted hairy black holes \cite{Ling:2020laa,Liu:2022pan,Geng:2025rov}. Specifically, how the emergence of scalar hair and the underlying tachyonic instability alter the fine structure of the Page curve, and whether the scalarization phase transition accelerates or delays the Page time.
    
    \item Second, while the present work focuses on tachyonic instability, it is crucial to explore how holographic quantum effects modify the physical bounds of other linear or nonlinear classical instabilities. For instance, in the context of superradiant instabilities in rotating or confined systems \cite{Bosch:2016vcp,East:2018glu,Chesler:2021ehz,Chen:2022vag}, the quantum-induced geometric deformation might either amplify the energy extraction process or completely quench it, potentially giving rise to novel quantum-corrected hairy configurations \cite{Cartwright:2025fay}.
    
    \item  Finally, our current stability analysis is restricted to static backgrounds and linear perturbations. Implementing the full real-time nonlinear dynamical evolution of the coupled bulk-brane system would completely unveil the time-dependent process of the quantum-enhanced scalarization. Such fully nonlinear simulations would not only determine the exact dynamical endpoints of the unstable configurations but also capture potential intermediate critical phenomena and dynamical barriers, offering a comprehensive picture of how macroscopic quantum effects govern black hole dynamics far from equilibrium.
\end{itemize}

%------------------------------------
%------------------------------------
\section*{Acknowledgments}
We are grateful to Yuan Sun, Qinghua Zhu, Zhuo-Yu Xian for the helpful discussions. Liu Yuxuan special thanks to Cao Peiwen and Liu Ziven for supporting his work.
Liu Yuxuan acknowledges the use of Gemini 3.1 Pro for language polishing and proofreading of the manuscript. Liu Yuxuan also acknowledges the assistance of OpenAI GPT-5.6 Sol with numerical-data analysis and manuscript revision. LYX is supported by the Natural Science Foundation of China under Grant No.~12405079, the Natural Science Foundation of Hunan Province, China (Grant No.~2025JJ60062), and Research start-up funds from the Central South University.
YL is supported by the Natural Science Foundation of China (Grant No. 12275275).
CQ is supported by the National Natural Science Foundation of China with Grant No. 12447129 and the fellowship from the China Postdoctoral Science
Foundation with Grant No. 2024M760691.
%------------------------------------
%------------------------------------
\appendix
\section{\texorpdfstring{Monitoring the norm of the DeTurck vector}{Monitoring the norm of the DeTurck vector}}
\label{app:deturck-convergence}

\begin{figure}[H]
    \centering
    \subfigure[]{\label{fig:deturck-convergence-curve}
    \includegraphics[width=0.42\linewidth]{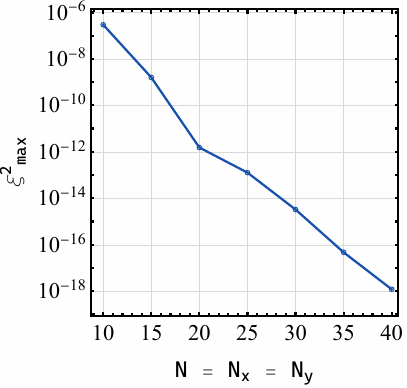}}
    \hfill
    \subfigure[]{\label{fig:deturck-distribution}
    \includegraphics[width=0.52\linewidth]{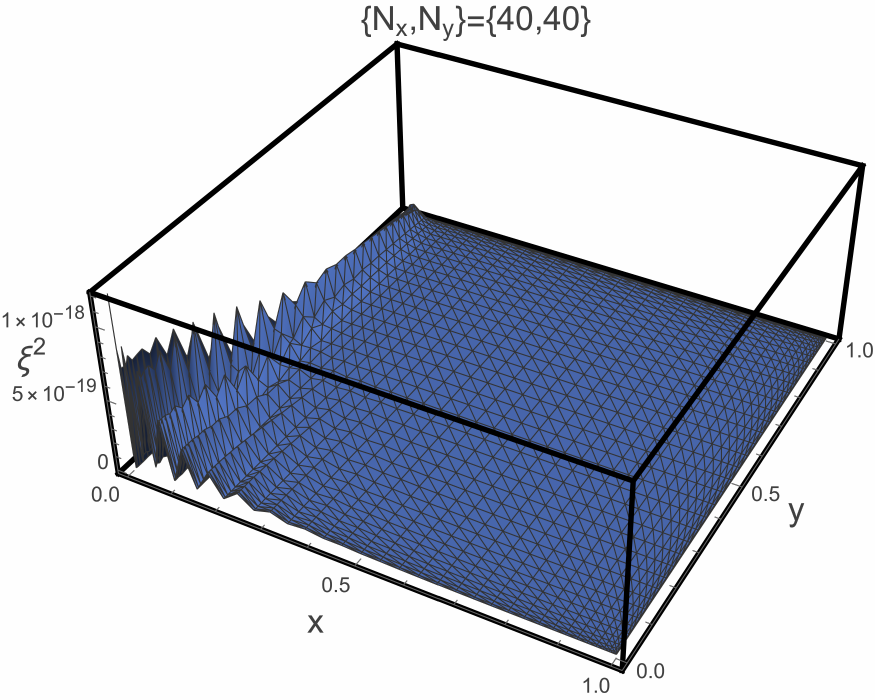}}
    \caption{Einstein--DeTurck convergence for the representative
    bald background specified in the text. (a) The maximum absolute value
    $\lVert\xi^2\rVert_\infty$ as a function of the equal grid size
    $N=N_x=N_y$. (b) The spatial distribution of $\xi^2$ at
    $N_x=N_y=40$.}
    \label{fig:deturck-convergence}
\end{figure}

We monitor the DeTurck residual through
\begin{equation}
    \lVert \xi^2\rVert_\infty
    :=\max_{(x,y)\in[0,1]^2}\left|\xi_\mu\xi^\mu\right| .
\end{equation}
As a representative test, we use the bald background with
\begin{equation}
    \{\alpha,\phi_s,\theta,y_h,L,\lambda,\mu\}
    =\{0,0,\pi/2,1,1,0,1\},\qquad
    \{\kappa,T_h/\mu\}=\{1,0.31831\}.
\end{equation}
We increase the number of Chebyshev collocation points from
$N_x=N_y=10$ to $40$. As shown in figure~\ref{fig:deturck-convergence},
$\lVert\xi^2\rVert_\infty$ decreases monotonically from
$2.96\times10^{-7}$ to $1.20\times10^{-18}$. At machine precision,
the smallest values should be interpreted as the residual floor rather
than as an independent estimate of the accuracy of the metric functions.

\section{\texorpdfstring{QNM convergence}{QNM convergence}}
\label{app:qnm-convergence}

\begin{figure}[H]
    \centering
    \includegraphics[width=0.48\linewidth]{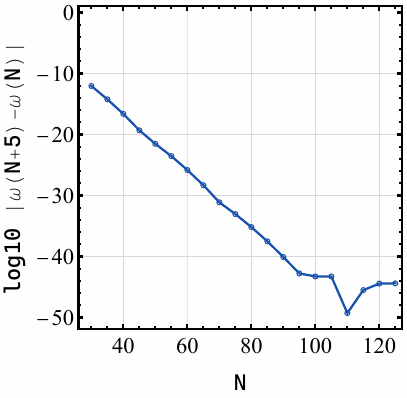}
    \caption{Convergence of a representative QNM frequency with the
    Chebyshev resolution $N$. The vertical axis shows
    $\log_{10}\left|\omega(N+5)-\omega(N)\right|$.}
    \label{fig:qnm-convergence}
\end{figure}

We test the convergence of the QNM eigenvalue problem on the
representative bald background
$\{\alpha,\theta,y_h,L,\mu\}=\{0,\pi/2,1,1,0\}$. Using three times machine precision,
we track the same QNM branch while increasing the number of Chebyshev
collocation points from $N=30$ to $130$. The plotted difference
\begin{equation}
    \epsilon_N=\left|\omega(N+5)-\omega(N)\right|
\end{equation}
decreases approximately exponentially over the resolved range and
then reaches the working-precision floor, as shown in
Figure~\ref{fig:qnm-convergence}.

\bibliographystyle{unsrt}

\bibliography{refs}

\end{document}